\documentclass{icrc2009}

\usepackage{amssymb}
\usepackage{graphicx}   
\usepackage[caption=false]{caption}    
\usepackage[font=footnotesize]{subfig} 
\usepackage{fixltx2e}
\usepackage{url}
\newcommand{\shorttitle}[1]%
{\markboth{Proceedings of the 31\MakeLowercase{$^{st}$} ICRC, {\L}\'{o}d\'{z} 2009}{#1} }
\newcommand{\etal}{\MakeLowercase{\textit{et al. }}} 

\begin{document}
\title{Searching for high-energy neutrinos in coincidence with gravitational
waves with the ANTARES and VIRGO/LIGO detectors}

\author{\IEEEauthorblockN{V\'eronique Van Elewyck\IEEEauthorrefmark{1},
			  for the ANTARES Collaboration\IEEEauthorrefmark{2}}
                            \\
\IEEEauthorblockA{\IEEEauthorrefmark{1} AstroParticule et Cosmologie (UMR 7164) \& Universit\'e Paris 7 Denis Diderot, \\Case 7020, F-75205 Paris Cedex 13, France \\
email: elewyck@apc.univ-paris7.fr}
\IEEEauthorblockA{\IEEEauthorrefmark{2} http://antares.in2p3.fr}}

\shorttitle{V. Van Elewyck \etal Coincident searches ANTARES/VIRGO-LIGO}
\maketitle

\begin{abstract}
 Cataclysmic cosmic events can be plausible sources of both
gravitational waves (GW) and high-energy neutrinos (HEN). Both GW and HEN are
alternative cosmic messengers that may escape very dense media and
travel unaffected over cosmological distances, carrying information
from the innermost regions of the astrophysical engines. For the same
reasons, such messengers could also reveal new, hidden sources that
were not observed by conventional photon astronomy.

Requiring the consistency between GW and HEN detection channels shall
enable new searches as one has significant additional information
about the common source. A neutrino telescope such as ANTARES can
determine accurately the time and direction of high energy neutrino
events, while a network of gravitational wave detectors such as LIGO
and VIRGO can also provide timing/directional information for
gravitational wave bursts. By combining the  information from these
totally independent detectors, one can search for cosmic events that
may arrive from common astrophysical sources.

  \end{abstract}


\section{Introduction}

Astroparticle physics has entered an exciting period with the recent development of experimental techniques that have opened new windows of observation of the cosmic radiation in all its components. In this context, it has been recognized that not only a multi-wavelength but also a multi-messenger approach was best suited for studying the high-energy astrophysical sources.

In this context, and despite their elusive nature, both high-energy neutrinos (HENs) and gravitational waves (GWs) are now  considered seriously as candidate cosmic messengers. Contrarily to high-energy photons (which are absorbed through interactions in the source and with the extragalactic background light) and cosmic rays (which are deflected by ambient magnetic fields, except at the highest energies), both HENs and GWs may indeed escape from dense astrophysical regions and travel over large distances without being absorbed, pointing back to their emitter.

It is expected that many astrophysical sources produce both GWs, originating from the cataclysmic event responsible for the onset of the source, and HENs, as a byproduct of the interactions of accelerated protons (and heavier nuclei) with ambient matter and radiation in the source. Moreover, some classes of astrophysical objects might be completely opaque to hadrons and photons, and observable only through their GW and/or HEN emissions. The detection of coincident signals in both these channels would then be a landmark event and sign the first observational evidence that GW and HEN originate from a common source. GW and HEN astronomies will then provide us with important information on the processes at work in the astrophysical accelerators. A more detailed discussion on the most plausible GW/HEN sources will be presented in Section \ref{sec:sources}, along with relevant references.

Furthermore, provided that the emission mechanisms are sufficiently well known, a precise measurement of the time delay between HEN and GW signals coming from a very distant source could also allow to probe quantum-gravity effects and possibly to constrain dark energy models\cite{QG}. Gamma-ray bursts (GRBs) appear as good candidates, {\it albeit} quite challenging\cite{QGlim}, for such time-of-flight studies.

Common HEN and GW astronomies are also motivated by the advent of a new generation of dedicated experiments. In this contribution, we describe in more detail the feasibility of joint GW+HEN searches between the ANTARES neutrino telescope\cite{antares} (and its future, $km^3$-sized, successor KM3NeT\cite{km3net}) and the GW detectors VIRGO\cite{virgo} and LIGO\cite{ligo} (which are now part of the same experimental collaboration). The detection principle and performances of these detectors are briefly described in Section \ref{sec:det}, along with their respective schedule of operation in the forthcoming years. Section \ref{sec:ana} presents an outlook on the analysis strategies that will be set up to optimize coincident GW-HEN detection among the three experiments.

It should be noted that similar studies involving the Ice Cube neutrino telescope\cite{ice3}, currently under deployment at the South Pole and looking for cosmic neutrinos from sources in the Northern Hemisphere, are under way\cite{aso}.

\begin{figure*}[!t]
   \centerline{{\includegraphics[width=5in]{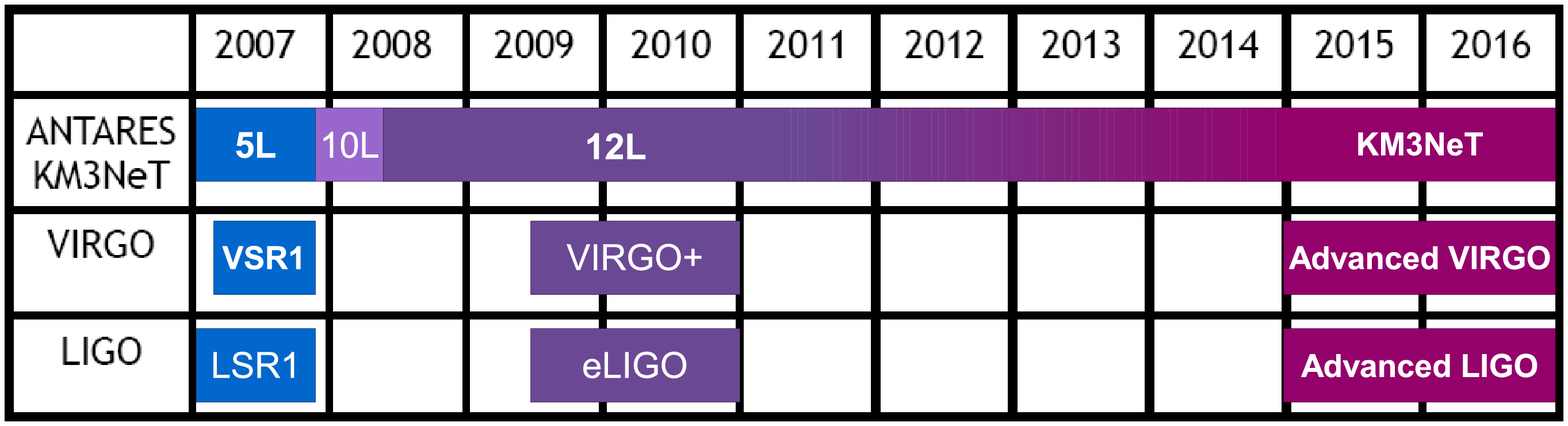} \label{sub_fig1}}}
\vspace*{-5.5cm}
   \caption{\footnotesize Time chart of the data-taking periods for the ANTARES, VIRGO and LIGO experiments, indicating the respective upgrades of the detectors (as described in the text). The deployment of the KM3NeT neutrino telescope is expected to last three to four years, during which the detector will be taking data with an increasing number of PMTs before reaching its final configuration.\cite{km3net}.}
   \label{fig:tab}
 \end{figure*}

\section{Potential common sources of Gravitational waves and high-energy neutrinos}
\label{sec:sources}
\subsection{Galactic sources}
 Several classes of galactic sources could feature both GW and HEN emission mechanisms potentially accessible to the present generation of GW interferometers and HEN telescopes, among which :
\begin{itemize}
 \item {\bf Microquasars} are associated with X-ray binaries involving a compact object (neutron star or black hole) that accretes matter from a companion star and re-emits it in relativistic jets associated with intense radio (and IR) flares. Such objects could emit GWs during both accretion and ejection phases\cite{pradierVLVNT}; and the latter phase could be correlated with a HEN signal if the jet has a hadronic component\cite{mq}.
\item {\bf Soft Gamma-ray Repeaters (SGRs)} are X-ray pulsars with a soft $\gamma$-ray bursting activity. The {\it magnetar} model described them as highly magnetized neutron stars whose outbursts are caused by global star-quakes associated with rearragements of the magnetic field\cite{magnetar}. The deformation of the star during the outburst could produce GWs, while HENs could emerge from hadron-loaded flares\cite{ioka}. The fact that both the AMANDA neutrino telescope and the LIGO interferometer have issued relevant limits on the respective GW and HEN emissions from the 2004 giant flare of SGR 1806-20\cite{sgrlim} indicates that a coincident GW+HEN search might further constrain the parameter space of emission models.
\end{itemize}

\subsection{Extragalactic sources}
Gamma-Ray Bursts (GRBs) are probably the most promising class of extragalactic sources. In the prompt and afterglow phases, HENs ($10^5 - 10^{10}$ GeV) are expected to be produced by accelerated protons in relativistic shocks and several models predict detectable fluxes in km$^3$-scale detectors\cite{GRBnus}.
\begin{itemize}
 \item {\bf Short-hard GRBs} are thought to originate from coalescing binaries involving  black holes and/or neutron stars; such mergers could emit GWs detectable from relatively large distances\cite{nakar}, with significant associated HEN fluxes.
\item {\bf Long-soft GRBs}, as described by the collapsar model, are compatible with the emission of a strong burst of GWs during the gravitational collapse of the (rapidly rotating) progenitor star and in the pre-GRB phase; however this population is distributed over cosmological distances so that the associated HEN signal is expected to be faint\cite{kotake}.
\item{\bf Low-luminosity GRBs}, with $\gamma$-ray luminosities a few orders of magnitude smaller, are believed to originate from a particularly energetic, possibly rapidly-rotating and jet-driven population of core-collapse supernovae. They could produce stronger GW signals together with significant high- and low-energy neutrino emission; moreover they are often discovered at shorter distances\cite{gupta}.
\item {\bf Failed GRBs} are thought to be associated with supernovae driven by mildly relativistic, baryon-rich and optically thick jets, so that no $\gamma$-rays escape. Such ``hidden sources'' could be among the most promising emitters of GWs and HENs, as current estimations predict a relatively high occurrence rate in the volume probed by current GW and HEN detectors\cite{ando}.
\end{itemize}

\section{The detectors}
 \label{sec:det}

\subsection{ANTARES}
  The ANTARES detector is the first undersea neutrino telescope; its deployment at a depth of 2475m in the Mediterranean Sea near Toulon was completed in May 2008. It consists in a three-dimensional array of 884 photomultiplier tubes (PMTs) distributed on 12 lines anchored to the sea bed and connected to the shore through an electro-optical cable. Before reaching this final (12L) setup, ANTARES has been operating in various configurations with increasing number of lines, from one to five (5L) and ten (10L); the respective periods are indicated on the time chart of Fig.~\ref{fig:tab}.
  
   ANTARES detects the Cherenkov radiation emitted by charged leptons (mainly muons, but also electrons and taus) induced by cosmic neutrino interactions with matter inside or near the instrumented volume. The knowledge of the timing and amplitude of the light pulses recorded by the PMTs allows to reconstruct the trajectory of the muon and to infer the arrival direction of the incident neutrino and to estimate its energy.
   
   Since the Earth acts as a shield against all particles but neutrinos, the design of a neutrino telescope is optimized for the detection of up-going muons produced by neutrinos which have traversed the Earth and interacted near the detector. The field of view of ANTARES is therefore $\sim\, 2 \pi\, \mathrm{sr}$ for neutrino energies $100\ \mathrm{GeV} \lesssim E_\nu \lesssim 100\  \mathrm{TeV}$. Above this energy, the sky coverage is reduced because of neutrino absorption in the Earth; but it can be partially recovered by looking for horizontal and downward-going neutrinos, which can be more easily identified as the background of atmospheric muons is much fainter at these energies. With an effective area $\sim 0,1\ \mathrm{km^2}$), ANTARES is expected to achieve an unprecedented angular resolution (about $0.3^\circ$ for neutrinos above 10 TeV) as a result of the good optical properties of sea water\cite{opt}.
  
  The data acquisition system of ANTARES is based on the "all-data-to-shore" concept, which allows to operate different physics triggers to the same data in parallel, each optimized for a specific (astro)physics signal\cite{miekedata}. In particular, satellites looking for GRBs can trigger the detector in real time via the GCN (Gamma-Ray Burst Coordinate Network) alert system. About 60s of buffered raw data (sometimes even including data recorded before the alert time) are then written on disk and kept for offline analysis, allowing for a significant gain in detection efficiency\cite{miekegrb}. 
  
  Another interesting feature recently implemented in ANTARES is the possibility to trigger an optical telescope network on the basis of "golden" neutrino events (such as neutrino doublets coincident in time and space or single neutrinos of very high energy) selected by a fast, online reconstruction procedure\cite{damien}.

   All these characteristics make the ANTARES detector especially suited for the search of astrophysical point sources\cite{simona} - and transients in particular.
  
\subsection{VIRGO and LIGO}

The GW detectors VIRGO\cite{virgo} (one site in Italy) and LIGO\cite{ligo} (two sites in the United States) are Michelson-type laser interferometers that consist of two light storage arms enclosed in vacuum tubes oriented at $90^\circ$ from each other. Suspended, highly reflective mirrors play the role of test masses. Current detectors are sensitive to relative displacements (hence GW amplitude) of the order of $10^{20}$ to $10^{22}$   Hz$^{-1/2}$. Their current detection horizon is about 15 Mpc for standard binary sources.

Their sensitivity is essentially limited at high frequencies by laser shot noise and at low frequencies by seismic noise ($<\ \sim O(50\ \mathrm{Hz})$) and by the thermal noise of the atoms in the mirrors (up to few 10 Hz). To reduce the sources of noise, GW interferometers rely on high-power, ultrastable lasers and on sophisticated techniques of position and alignment control and of stabilization of the mirrors.

Both detectors had a data-taking phase during 2007 (Virgo Science Run 1 and Ligo S5), which partially coincided with the ANTARES 5L configuration. They are currently upgrading to VIRGO+ and eLIGO, to improve their sensitivity by a factor of 2 - and hence the probed volume by a factor of 8. The collaborations have merged and are preparing for a common science run starting mid-2009 (Virgo Science Run 2 and LIGO S6), i.e. in coincidence with the operation of ANTARES 12L (see Fig.~\ref{fig:tab}). Another major upgrade for both classes of detectors is scheduled for the upcoming decade: the Advanced VIRGO/Advanced LIGO and KM3NeT projects should gain a factor of 10 in sensitivities respect to the presently operating instruments. 

\begin{figure}[!t]
  \centering
  \includegraphics[width=3in]{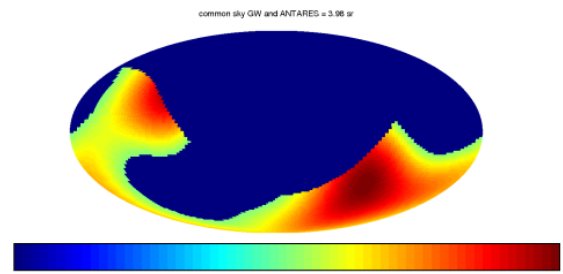}
  \caption{\footnotesize Instantaneous common sky coverage for VIRGO + LIGO + ANTARES in geocentric coordinates. This map shows the combined antenna pattern for the gravitational wave detector network (above half-maximum), with the simplifying assumption that ANTARES has 100\% visibility in its antipodal hemisphere and 0\% elsewhere. The colour scale is from $0\%$ (left, blue) to $100\%$ (right, red)\cite{ecm}.}
  \label{fig:map}
 \end{figure}

The VIRGO/LIGO network monitors a good fraction of the sky in common with HEN telescopes: as can be seen from Figure~\ref{fig:map} the overlap of visibility maps with each telescope is about 4~sr ($\sim 30\%$ of the sky), and the same value holds for Ice Cube.

\section{Outlook on the analysis strategies}
\label{sec:ana}

GW interferometers and HEN telescopes share the challenge to look for faint and rare signals on top of abundant noise or background events. The GW+HEN search methodology involves the combination of independent GW/HEN candidate event lists, with the advantage of significantly lowering the rate of accidental coincidences.

The information required about any GW/HEN event consists of its timing, arrival direction and associated angular uncertainties (possibly under the form of a sigificance sky map). Each event list is obtained by the combination of  reconstruction algorithms specific to each experiment, and quality cuts used to optimize the signal-to-background ratio.

GW+HEN event pairs within a predefined, astrophysically motivated (and possibly source- or model-dependent), time interval can be selected as time-coincident events. Then, the spatial overlap between GW and HEN events is statistically evaluated, e. g. by an unbinned maximum likelihood method, and the significance of the coincident event is obtained by comparing to the distribution of accidental events obtained with Monte-Carlo simulations using data streams scrambled in time (or simulated background events).

Preliminary investigations of the feasibility of such searches have already been performed and indicate that, even if the constituent observatories provide several triggers a day, the false alarm rate for the combined detector network can be maintained at a very low level ($\sim (600\ \mathrm{yr})^{-1}$)\cite{aso,pradierVLVNT}.

\section{Conclusions and perspectives}
\label{sec:concl}

A joint GW+HEN analysis program could significantly expand the scientific reach of both GW interferometers and HEN telescopes. The robust background rejection arising from the combination of two totally independent sets of data results in an increased sensitivity and the possible recovery of cosmic signals. The observation of coincident triggers would provide strong evidence for the detection of a GW burst and a cosmic neutrino event, and for the existence of common sources. Beyond the benefit of a high-confidence discovery, coincident GW/HEN (non-)ob\-servation shall play a critical role in our understanding of the most energetic sources of cosmic radiation and in constraining existing models. They could also reveal new, ``hidden'' sources unobserved so far by conventional photon astronomy.
A new period of concurrent observations with upgraded experiments is expected to start mid-2009. Future schedules involving next-generation detectors with a significantly increased sensitivity (such as KM3NeT and the Advanced LIGO/Advanced VIRGO projects) are likely to coincide as well, opening the way towards an even more efficient GW+HEN astronomy.

\section{Acknowledgements}
 V. V. E.  thanks E. Chassande-Mottin for many fruitful discussions and for his help in preparing this manuscript. She acknowledges financial support from the European Community 7th Framework Program (Marie Curie Reintegration Grant) and from the French Agence Nationale de la Recherche (ANR-08-JCJC-0061-01).

\end{document}